\documentclass[aps,pra,reprint,longbibliography]{revtex4-1}
\usepackage{graphicx,color,transparent}
\usepackage{amsmath,braket}
\usepackage{lipsum,hyperref,layouts,float}

\newcommand{\dm}{\vert\langle 5P_{3/2}\vert\vert er\vert\vert 5D_{5/2}\rangle\vert}
\newcommand{\dmreduced}{\vert\langle 5P\vert\vert er\vert\vert 5D\rangle\vert}
\newcommand{\rb}{$^{87}$Rb}

\begin{document}
\title{Direct measurement of excited-state dipole matrix elements using electromagnetically induced transparency in the hyperfine Paschen-Back regime}
\author{Daniel J Whiting}
\email{daniel.whiting@durham.ac.uk}
\author{James Keaveney, Charles S Adams, Ifan G Hughes}
\affiliation{Department of Physics, Joint Quantum Centre (JQC) Durham-Newcastle, Rochester Building, Durham University, South Road, Durham, DH1 3LE, UK}
\date{\today}
\begin{abstract}
Applying large magnetic fields to gain access to the hyperfine Paschen-Back regime can isolate three-level systems in a hot alkali metal vapors, thereby simplifying usually complex atom-light interactions.
We use this method to make the first direct measurement of the $\dmreduced$ matrix element in \rb.
An analytic model with only three levels accurately models the experimental electromagnetically induced transparency spectra and extracted Rabi frequencies are used to determine the dipole matrix element.
We measure $\dm = (2.290\pm0.002_{\rm stat}\pm0.04_{\rm syst})~ea_{0}$ which is in excellent agreement with the theoretical calculations of Safronova, Williams, and Clark [Phys. Rev. A \textbf{69}, 022509 (2004)].
\end{abstract}
\maketitle

\section{Introduction}
Understanding coherent atom-light interactions in multi-level atomic media continues to be the focus of much attention, with an increasing variety of applications.
Phenomena such as electromagnetically induced transparency (EIT)~\cite{Harris1997,Fleischhauer2005}, coherent population trapping (CPT)~\cite{Gray1978}, and four-wave mixing (FWM) have been instrumental in the development of atomic clocks~\cite{Hong2005,Santra2005}, magnetometers~\cite{Yudin2010}, sub- and super-luminal propagation of light~\cite{Hau1999,Keaveney2012}, quantum memories~\cite{Lukin2000}, photonic phase gates~\cite{Friedler2005}, single-photon sources~\cite{MacRae2012a,Srivathsan2013}, and squeezed states of light~\cite{Marino2008}.

Quantitative modeling for thermal ensembles remains difficult due to the complex atomic energy-level structure.
In contrast, there has been much recent progress in quantitative modeling of linear systems both in terms of fundamental physics and applications \cite{Keaveney2012,Weller2012,Weller2012a,Zentile2014,Zentile2015a,Zentile2015}.
A recent experiment \cite{Zentile2014} demonstrated that applying a large magnetic field can greatly simplify spectroscopic measurements, allowing the isolation of pure two-level systems even in hot atomic vapors where the Doppler broadening is significantly larger than the natural linewidth.
Here we show this can be extended to two-photon excitation which, in a similar way, isolates pure three-level systems and thereby allows accurate modeling of EIT spectra.
As a demonstration we use this method to make the first direct measurement of the $\dmreduced$ matrix element in \rb.

\begin{figure}[!hb]
\includegraphics[width=0.8\columnwidth]{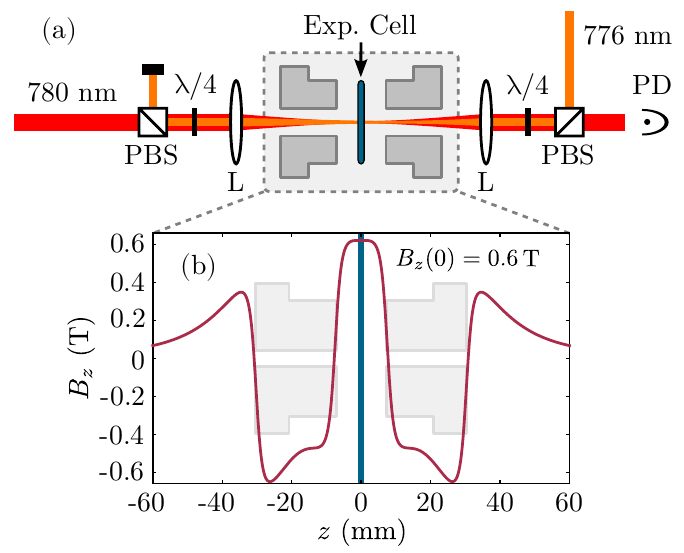}
\caption{\label{fig:setup}
		(a) Schematic of the experimental apparatus. PBS - Polarizing beam splitter; $\lambda/4$ - quarter waveplate; L - lens with 200 mm focal length; Exp. Cell - 2~mm rubidium vapor cell; PD - high gain photodetector.
		(b) Axial magnetic field profile (red) of the permanent ring-magnets (grey blocks). The maximum field strength is 0.6~T and the maximum variation over the region occupied by the vapor cell (shown in blue) is less than 1~mT.}
\end{figure}

\begin{figure*}[t]
\includegraphics[width=\textwidth]{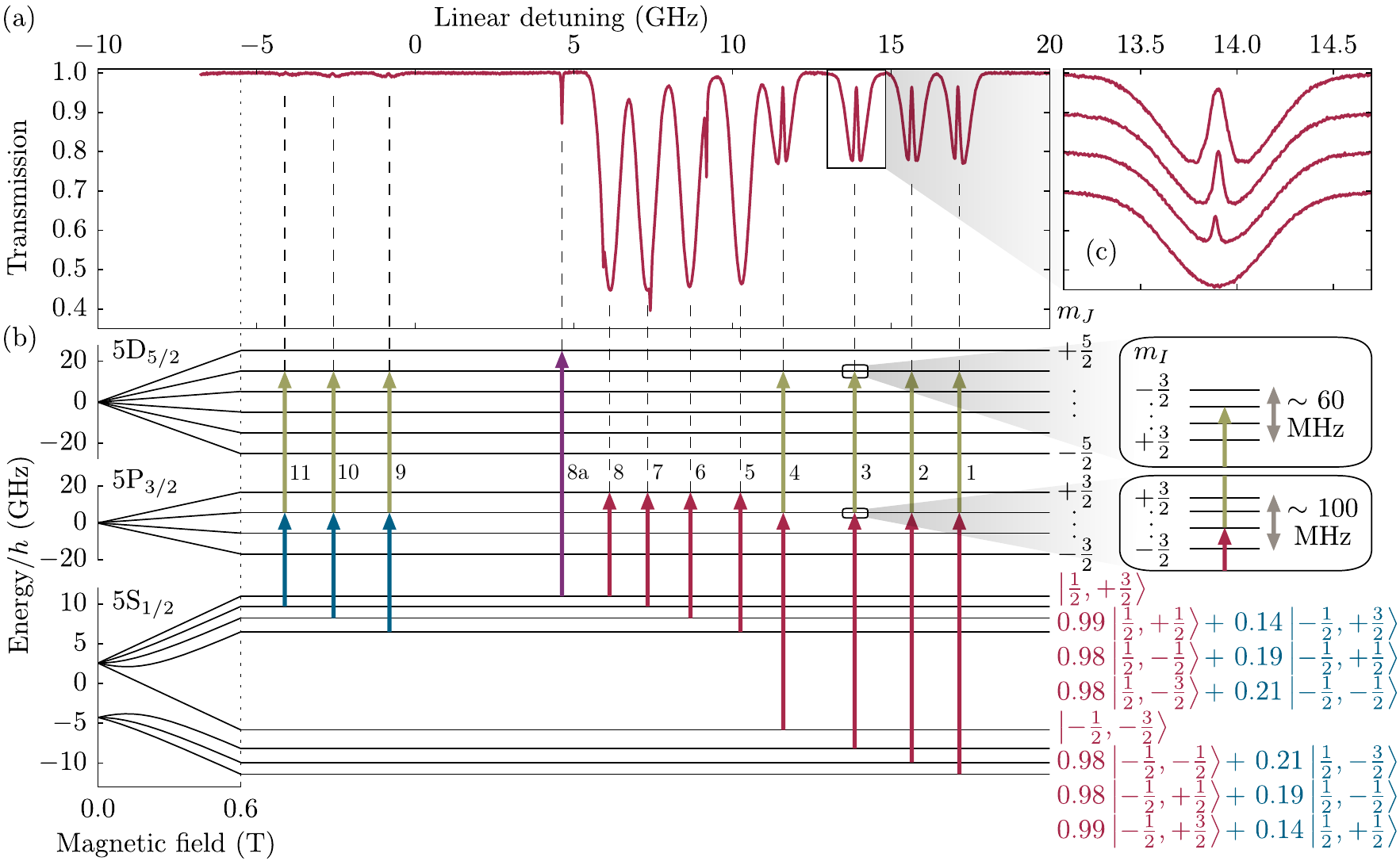}
\caption{\label{fig:BRdiagram}
		Panel (a) shows a typical weak probe transmission spectrum (red).
		Panel (b) is a diagram of the transitions associated with each of the spectral features in panel (a).
		On the left of the panel is an energy level diagram for the relevant states, showing the evolution into the hyperfine Paschen-Back regime at large magnetic fields.
		The eigenstates of the system in the $\ket{m_{J},m_{I}}$ basis are shown to the right of the diagram for a magnetic field strength $B=0.6$~T.
		Even at this large field there still exists a significant admixture of states with opposite spin (blue text) in the 5S$_{1/2}$ manifold, these result in the weak transitions indicated by the blue arrows.
		The large splittings between ground states as well as the electric dipole selection rule $\Delta m_{I}=0$ ensure that only three atomic states are involved in each of the two-photon resonances.
		An expanded view of the three-level EIT resonance $5S_{1/2}\ket{-\frac{1}{2},-\frac{1}{2}}\rightarrow 5P_{3/2}\ket{\frac{1}{2},-\frac{1}{2}}\rightarrow 5D_{5/2}\ket{\frac{3}{2},-\frac{1}{2}}$ is shown in panel (c) for control-beam powers of 0, 3, 9 and 27~mW. Vertical offsets have been added for clarity.
		Zero detuning is given as the weighted $D_{2}$ line center of naturally abundant rubidium in zero-magnetic field~\cite{Siddons2008}.}
\end{figure*}

Matrix elements for the alkali-metal atom $D$-lines are known to a very high precision by measuring fluorescence decay lifetimes \cite{Volz1996}. 
Extending this technique to transitions between excited states is generally very difficult owing to the many possible decay channels.
In most cases these matrix elements can be inferred by combining experimentally measured lifetimes with theoretically predicted branching ratios.
However, in the case of $D$ states, properties such as lifetimes and branching ratios are more difficult to calculate accurately \cite{Sahoo2016,Safronova2004}.
Direct experimental measurements of these matrix elements are therefore needed to support theoretical models.

In this article we study EIT in the hyperfine Paschen-Back (HPB) regime, allowing us to isolate a pure three-level system in a hot atomic vapor of \rb.
Previous work on EIT in large magnetic fields focused on $\Lambda$-EIT~\cite{Sargsyan2012} and CPT~\cite{Holler1997} and did not include quantitative analysis of the spectra.
We extend this work by investigating a ladder system, which is of interest in for example, Rydberg EIT~\cite{Mohapatra2007} and diamond four-wave mixing~\cite{Kolle2012,Willis2009}.
Following the method of \cite{Piotrowicz2011} we conduct quantitative analysis of the EIT spectra to determine Rabi frequencies and extract the dipole matrix element.
We find excellent agreement with an analytic three-level model for the spectra \cite{Gea-Banacloche1995} giving confidence in the reported matrix element.

\section{Experimental details}
A schematic diagram of the experimental apparatus is shown in Fig.~\ref{fig:setup}.
We use a 2~mm long vapor cell, containing isotopically enriched \rb ~($<$2\% $^{85}$Rb), which is heated to 70~$^{\circ}$C to provide suitable optical depth on resonance.
The cell also contains a buffer gas which leads to a Lorentzian line broadening of $\sim$7~MHz on the 5S$\rightarrow$5P transitions and $\sim$20~MHz on the 5P$\rightarrow$5D transitions ~\cite{Sargsyan2010}.
To generate the strong magnetic field needed to study the HPB regime we use a pair of permanent magnets separated by 15~mm.
A cross sectional view is shown in Fig.~\ref{fig:setup}(b) along with the axial magnetic field profile.
The field strength at the center of the magnets is 0.6~T and has a variation of less than 1~mT over the length of the vapor cell (Blue line in Fig.~\ref{fig:setup}).
We measure the transmission of a weak~\cite{Sherlock2009} (20~nW) probe laser at 780~nm as its frequency is scanned $\sim$25~GHz around the Rubidium $D_{2}$ ($5S_{1/2}\rightarrow5P_{3/2}$) transition.
A second ECDL at 776~nm provides a strong control field resonant with the $5P_{3/2}\rightarrow5D_{5/2}$ transition which counter-propagates with the first.
The beams are focused through the cell to $1/\mathrm{e}^{2}$ radii of $54\pm1~\mu$m for the probe beam and $115\pm1~\mu$m for the control beam.
The beam waists are chosen as a compromise between maximizing beam intensity, minimizing beam diffraction across the cell and ensuring that the probed atoms observe a uniform control-beam electric field profile.
The polarizations are set using $\lambda/4$ waveplates such that they drive $\sigma^{+}$ ($\Delta m_{J}=+1$) transitions.
The temperature and magnetic field strength are verified by quantitatively fitting the probe transmission spectra with the control beam blocked~\cite{Zentile2015}.

\section{Electromagnetically induced transparency spectra}
An example transmission spectrum is shown in Fig.~\ref{fig:BRdiagram}(a).
The spectroscopic features can be mapped on to individual $\Delta m_{J} = +1$ transitions, as we show in Fig.~\ref{fig:BRdiagram}(b).
To understand the probe transmission spectrum we first consider the effect of the strong magnetic field on the energy levels of the atoms.
The magnetic field splits the energy levels according to their magnetic spin-orbit quantum numbers ($m_{J}$), with the change in energy ${\Delta E\approx m_{J}\mu_{\rm B}B}$~\cite{Bransden2000}, where $\mu_{\rm B}$ is the Bohr magneton and $B$ is the (axial) magnetic field strength.

For \rb ~($I=3/2$) in a 0.6~T magnetic field the Zeeman shift is much larger than the hyperfine interaction energy.
This leads to a decoupling of the spin-orbit and nuclear spin angular momenta which defines the hyperfine Paschen-Back regime \cite{Bransden2000}.
In the $5S_{1/2}$ ground state the hyperfine structure is split into two groups of states with magnetic spin-orbit quantum numbers $m_{J}=\pm1/2$, each of which contain $2I+1=4$ states with $m_{I}=\pm1/2,\pm3/2$ that are split by the hyperfine interaction energy.
In the excited states $J$, the spin-orbit angular momentum, is larger so that both the number of states and the splitting of the stretched states increase.
The hyperfine interaction energy is largest in the ground state where it is also larger than the Doppler width.
Coupled with the transition selection rule $\Delta m_{I}=0$ this means that the probe field can address a single $m_{I},~m_{J}\rightarrow m_{J}+1$ component of the $5S_{1/2}$ to $5P_{3/2}$ transition.

At 0.6~T the interaction of the ground state with the field is not large enough to completely decouple the spin-orbit and nuclear spin angular momenta.
On the right hand side of Fig.~\ref{fig:BRdiagram} we write the $5S_{1/2}$ eigenstates in the uncoupled basis, $|m_{J},m_{I}\rangle$, which show clearly (in blue) the small fraction of residual hyperfine mixing.
This small mixed fraction gives rise to the three weak transitions labeled 9-11 on Fig.~\ref{fig:BRdiagram}(b).
These weak transitions and the dependence of their line strengths on magnetic field have been investigated in detail in~\cite{Sargsyan2014}.

In our experiment the control field is resonant with the $5P_{3/2}\ket{\frac{1}{2},-\frac{1}{2}}\rightarrow5D_{5/2}\ket{\frac{3}{2},-\frac{1}{2}}$ transition.
This leads to Doppler-free EIT resonances on each of the lines labeled 1-4 and 9-11 as well as some narrow off-resonant two-photon absorption lines associated with probe absorption resonances labeled 5-8.
For clarity one of these two-photon absorption resonances is indicated by an additional arrow labeled 8a.

\begin{figure}[t]
\includegraphics[width=\columnwidth]{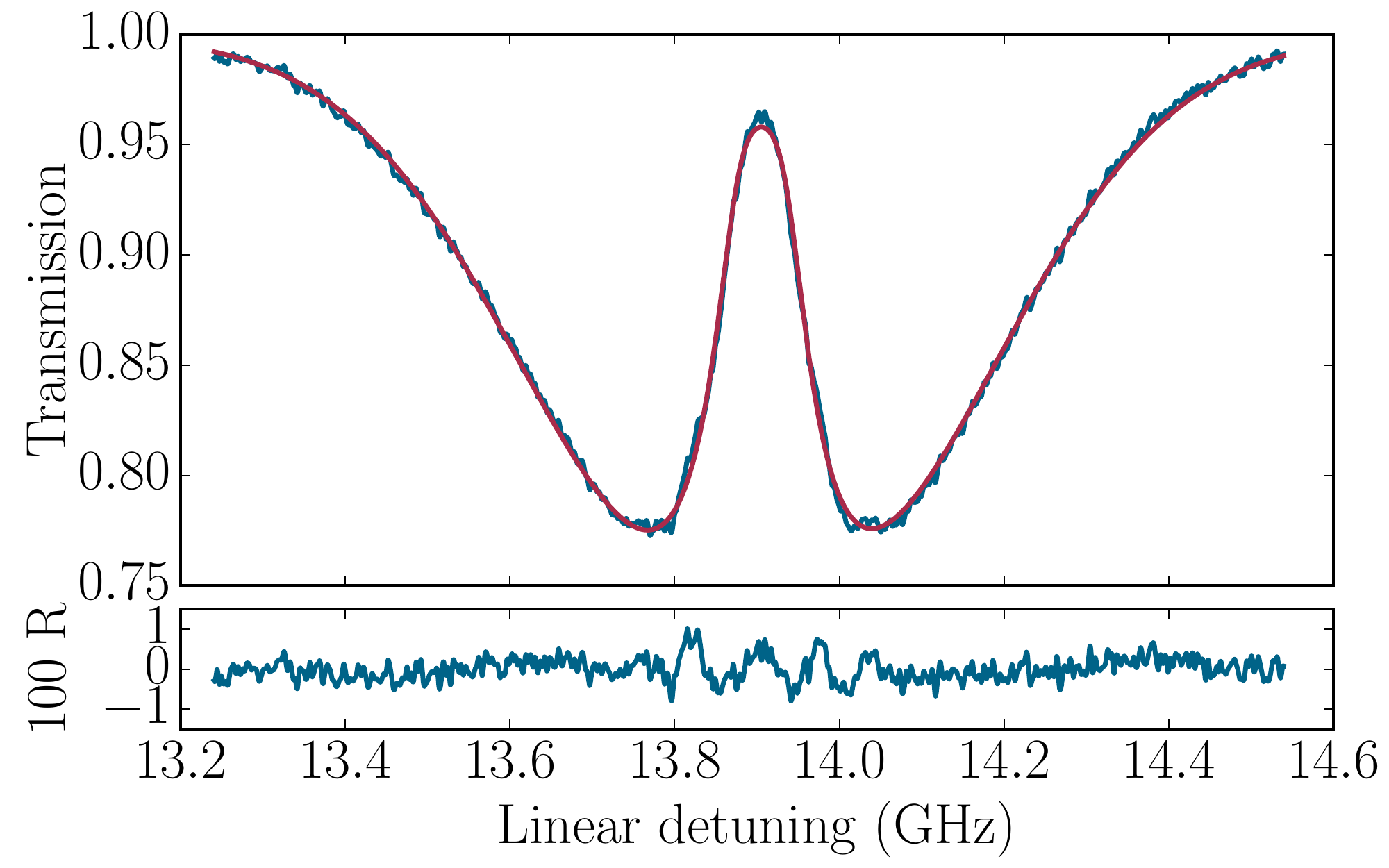}
\caption{\label{fig:single-fit} Experimental transmission spectrum (blue) showing a purely three-level EIT resonance in a hot $^{87}$Rb vapor.
		 The red line is a least-squares fit to the data using a three-level EIT model. The fit results in a measurement of the control-beam Rabi frequency $\Omega_{\rm c}/2\pi = 255.0\pm0.2$~MHz where the uncertainty is from the statistical error of the fit.
		 The residual (R) shows the excellent agreement between the model and data, with the small amount of structure near line center being explained by electromagnetically induced absorption (see main text).}
\end{figure}

Crucially, all these two-photon resonances involve only 3 atomic states; this is in contrast to the zero-field case when it is typically necessary to include large numbers of degenerate states to correctly model the response of the medium~\cite{Badger2001}.
We now focus on a single absorption line and show, in Fig.~\ref{fig:single-fit}, that we are able to accurately predict the observed EIT line shapes with a simple three-level model.
The electric susceptibility of a medium containing three level ladder-type atoms (with axial velocity component $v$) exposed to a strong control field and a weak probe field is given by the formula~\cite{Gea-Banacloche1995},
\begin{equation}\label{equation:eit}
\chi(v) = \frac{4i\hbar d_{\mathrm{p}}^{2}/\epsilon_{0}}{\gamma_{\mathrm{p}}-i\delta_{1}(v)+\displaystyle\frac{\Omega_{\mathrm{c}}^{2}/4}{\gamma_{\mathrm{c}}-i\delta_{2}(v)}}N,
\end{equation}
where $\delta_{1}(v)=\Delta_{\mathrm{p}}+k_{\mathrm{p}}v$, $\delta_{2}(v)=\Delta_{\mathrm{p}}+\Delta_{\mathrm{c}}+(k_{\mathrm{p}}-k_{\mathrm{c}})v$, $d_{\rm p}$~is the dipole matrix element of the probe transition, $\epsilon_{0}$ is the vacuum permittivity, $\gamma_{\mathrm{p,c}}$ are the decay rates of the coherence on the probe and control transitions, $\Delta_{\mathrm{p,c}}$ are the probe and control detunings, $k_{\mathrm{p,c}}$ are the probe and control wavenumbers, $\Omega_{\mathrm{c}}$ is the control transition Rabi frequency and $N$ is the number density of $^{87}$Rb atoms.
The absorption coefficient is then given by the integral of the imaginary component of the susceptibility over the Maxwell-Boltzmann distribution of atomic velocities $n(v)=\mathrm{exp}(-v^{2}/u^{2})/(u\sqrt{\pi})$, where $u=\sqrt{2k_{\rm B}T/m}$ is the most probable speed, $k_{\rm B}$ is the Boltzmann constant, $T$ is the vapor temperature, and $m$ is the atomic mass.
The transmission of the weak probe field for a medium of length $l$ is given by
\begin{equation}\label{eqn:transmission}
T = \exp\left(-k_{\mathrm{p}}l\int_{-\infty}^{\infty}\mathrm{Im}[\chi(v)]n(v)~dv\right).
\end{equation}

Figure~\ref{fig:single-fit} shows experimental transmission spectra as well as a least-squares fit to the above model.
It is clear that the model is in excellent agreement with the data.
However, there is some small but noticeable structure near to line center which is consistent with a small amount of electromagnetically induced absorption~\cite{Whiting2015} caused by the back-reflected control light in the cell (see Appendix~\ref{EIA}).
We point out that such good agreement is not obtained for the case of zero magnetic field where multiple overlapping resonances lead to a more complex line shape, the modeling of which is non-trivial.

\begin{figure}
\includegraphics[width=\columnwidth]{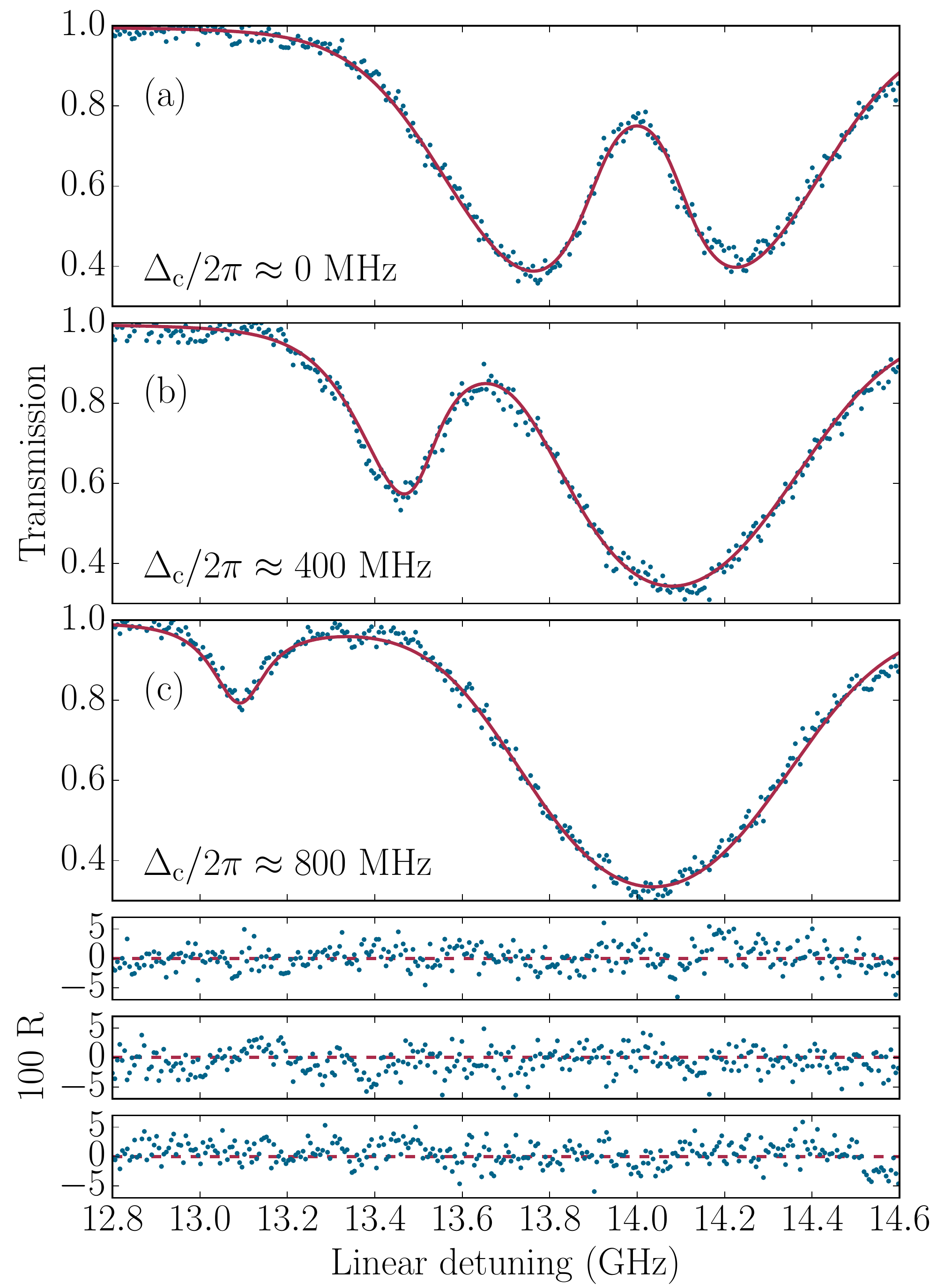}
\caption{\label{fig:detuning}
		Experimental transmission spectra (blue) and numerical fits (red) for three control-beam detunings.
The fit parameters except for control detuning are constrained to be equal in all three data sets.
The extracted control-beam detunings are $\Delta_{\mathrm{c}}/2\pi=-18.8\pm 0.6,~408.0\pm 0.6,~834.7 \pm 0.7$~MHz. The residuals (R) show the excellent agreement between the model and data.}
\end{figure}

Figure~\ref{fig:detuning} shows experimental spectra and fits for three different control detunings, $\Delta_{\mathrm{c}}/2\pi \approx 0,400,800$~MHz.
The transition from resonant EIT to off-resonant two-photon absorption is clearly displayed across this range providing an excellent test for the model.
The fits are to the three level EIT model [Eq.~(\ref{equation:eit})] and all fit parameters, except for control detuning, are constrained to be equal for all data sets.
The resulting fits are all excellent as shown by the structureless residuals~\cite{Hughes2010}.

\vfill\eject
\section{Dipole matrix element measurement}
\begin{figure}
\includegraphics[width=\columnwidth]{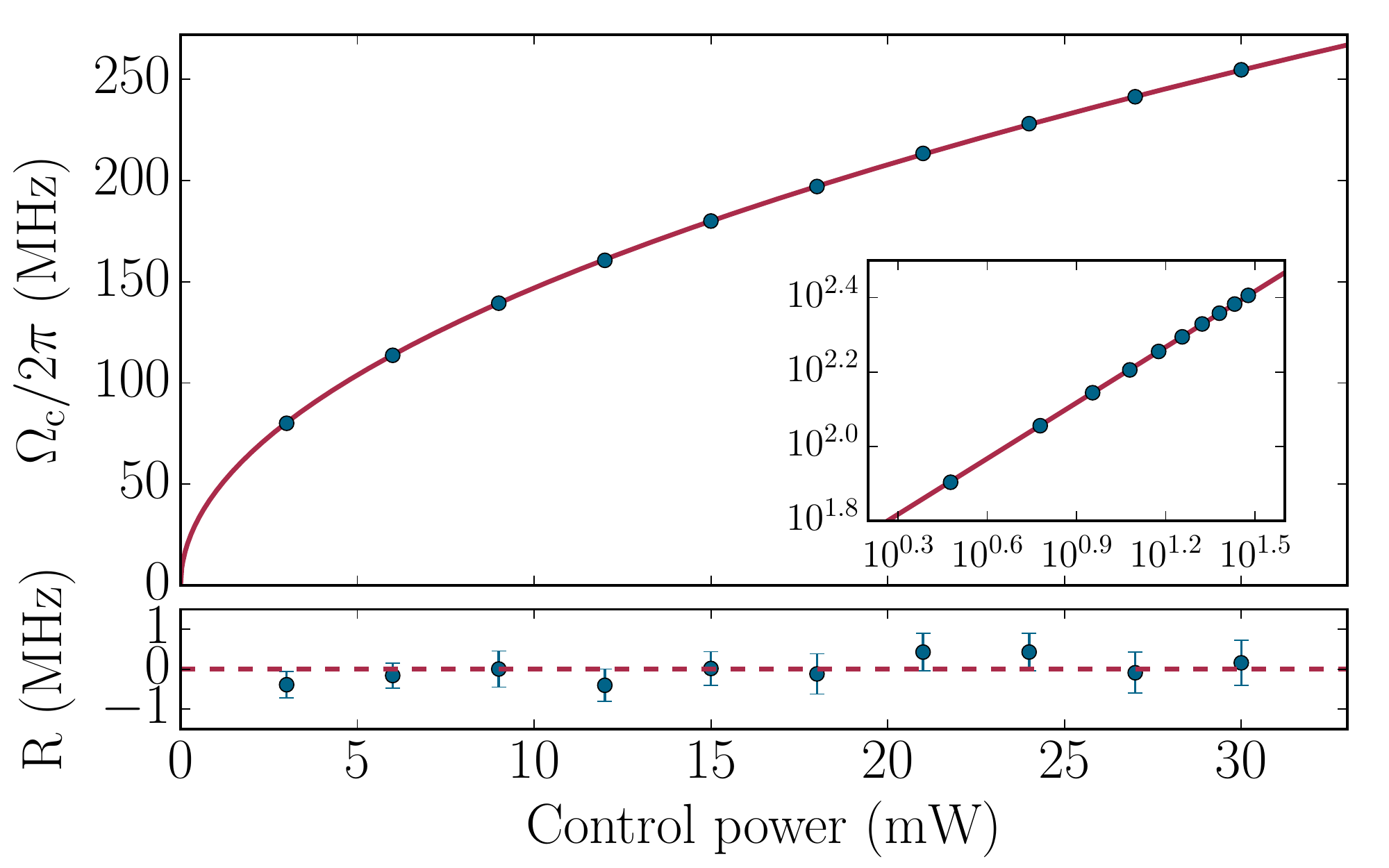}
\caption{\label{fig:rabi}
		Extracted values of the control Rabi frequency ($\Omega_{c}$) for increasing control-beam power ($P$).
		The red line is a least-squares fit to the function $\Omega_{\rm c}=\alpha\sqrt{P}$, with $\alpha/2\pi=(46.47\pm0.04)$~MHz$/\sqrt{\rm mW}$.
		The residuals (R) are shown below and the fit has a reduced chi-squared of 0.5. 
		The inset shows the data on a logarithmic scale.}
\end{figure}

Based on the excellent agreement between experiment and theory, we now extract the control Rabi frequency from the numerical fitting, for a range of control-beam powers and detunings, allowing us to measure the dipole matrix element $\dm$.
At each power Rabi frequencies are extracted for 15 control-beam detunings (between $-80$ and $+80$ MHz) and are averaged to give the data points shown in Fig.~\ref{fig:rabi}; the error bars shown are the standard error on the mean.
Since the Rabi frequency is proportional to the local electric field amplitude, we expect to observe a square root dependence on control-beam power.
In Fig.~\ref{fig:rabi} the red line is the result of a least-squares fit to the experimental data using the function $\Omega_{\rm c}=\alpha\sqrt{P}$, from which it is determined $\alpha/2\pi=(46.47\pm0.04)$~MHz$/\sqrt{\rm mW}$.
The agreement between model and experimental data is excellent, with structureless residuals and reduced chi squared $\chi_{\nu}^{2}=0.5$~\cite{Hughes2010}.
In the following section we provide details of how $\alpha$ is converted into the dipole matrix element.

\vfill\eject
\subsection{Dipole matrix element calculation}\label{angmom}
To calculate the dipole moment we must determine the control-beam electric field in the vicinity of the atoms.
Since the control-beam intensity changes significantly over the extent of the probe beam, we use a weighted average of the control-beam electric field profile ($E_{\rm control}$) weighted by the probe-beam intensity profile ($I_{\rm probe}$), i.e.
\begin{align*}
E_{0} = \frac{\int\int I_{\rm probe}(x,y)E_{\rm control}(x,y)~dxdy}{\int\int I_{\rm probe}(x,y)~dxdy}.
\end{align*}
We then determine the dipole moment of the driven transition (between ground and excited states $\ket{g}$ and $\ket{e}$) through the equation:
\begin{align*}
|\langle g|er_{+1}|e\rangle| &= \hbar\frac{\Omega_{\rm c}/\sqrt{P}}{E_{0}/\sqrt{P}} \\
&= (0.7110\pm0.0006_{\rm stat}\pm0.01_{\rm syst})~ea_{0}
\end{align*}
where the +1 refers to the fact that the transition is a $\sigma^{+}$ transition in which the final magnetic spin-orbit angular momentum $m_{J}^{\prime}=m_{J}+1$.
In a magnetic field of 0.60~T the states resonantly coupled by the control field can be decomposed in the $\ket{m_{L},m_{S}}$ basis as
\begin{align*}
&\ket{g} = 0.817\ket{0,\frac{1}{2}}+0.577\ket{1,-\frac{1}{2}}\\
&\ket{e} = 0.910\ket{1,\frac{1}{2}}+0.415\ket{2,-\frac{1}{2}}.
\end{align*}
Therefore,
\begin{align*}
&|\langle g|er_{+1}|e\rangle|\\
&= 0.982|\langle5P_{3/2},~m_{J}=1/2|er_{+1}|5D_{5/2},~m_{J}=3/2\rangle|.
\end{align*}
We now calculate the reduced matrix element using \cite{Weissbluth1978}
\begin{align*}
|\langle J,m_{J}|er_{+1}|J^{\prime},m_{J^{\prime}}\rangle| = 
\begin{pmatrix}
J & 1 & J^{\prime} \\
-m_{J} & +1 & m_{J^{\prime}}
\end{pmatrix}
|\langle J||er||J^{\prime}\rangle|
\end{align*}
which gives
\begin{align*}
&|\langle5P_{3/2}||er||5D_{5/2}\rangle| \\
&= \sqrt{10} |\langle5P_{3/2},~m_{J}=1/2|er_{+1}|5D_{5/2},~m_{J}=3/2\rangle|\\
&= (2.290\pm0.002_{\rm stat}\pm0.04_{\rm syst})~ea_{0}.
\end{align*}
which is in excellent agreement with the theoretically calculated value of $2.334~ea_{0}$~\cite{Safronova2004}.
Further, we may determine the fully reduced dipole matrix element using \cite{Weissbluth1978}
\begin{align*}
|\langle J||er||J^{\prime}\rangle| = \sqrt{(2J+1)(2J^{\prime}+1)}
\begin{Bmatrix}
J^{\prime} & 1 & J \\
L & S & L^{\prime}
\end{Bmatrix}
|\langle L||er||L^{\prime}\rangle|
\end{align*}
which gives 
\begin{align*}
&|\langle5P||er||5D\rangle| = \frac{5}{\sqrt{30}}|\langle5P_{3/2}||er||5D_{5/2}\rangle| \\
&= (2.10\pm0.002_{\rm stat}\pm0.04_{\rm syst})~ea_{0}.
\end{align*}

\vfill\eject
\subsection{Uncertainties}
Table \ref{tab:errors} shows a breakdown of the uncertainties in the measurement.
\begin{table}[t]
\begin{tabular}{ccc}
\hline\hline
Source & Correction (\%) & Uncertainty (\%) \\
\hline
Statistical & & 0.08 \\
Optical power meter & & 1.5 \\
Beam spatial profiles & & 0.7 \\
Control polarization purity & 0.5 & 0.05 \\
Vapor cell transmittance & 8.2 & 0.3 \\
Line shape systematics & & 0.5 \\
Total & & 2 \\
\hline\hline
\end{tabular}
\caption{\label{tab:errors} Error budget of the $\dm$ dipole moment measurement.
		The second column shows the corrections we have made to account for the measured polarization impurity of the control field and for the measured reflectivity of the vapor cell windows.}
\end{table}
The major sources of systematic uncertainty are the optical power meter (Thorlabs S121C sensor) and the beam profile measurements.
Optical power meters for these powers typically have calibration uncertainties of 3\% or more, therefore this method could potentially be used to more precisely calibrate such devices.
To reduce the systematic uncertainty to less than 1\% it becomes necessary to consider many sources of systematic changes to the line shape such as the small amount of electromagnetically induced absorption ~\cite{Whiting2015} we observe in the transmission spectra.
A breakdown of these systematic line shape effects is given in Appendix~\ref{lineshapesystematics}.

\section{conclusion}
In conclusion, we have demonstrated that the HPB regime vastly reduces the complexity of modeling coherent atom-light interactions in thermal vapors. 
We have used this technique to directly measure an excited-state dipole matrix element, which up to now has only been possible indirectly through lifetime measurements. 
This simple approach can be applied to many different systems, opening up new possibilities for studying coherent phenomena in many-level systems such as EIT/EIA, FWM, and quantum memories.

\section{acknowledgments}
The authors would like to thank E. Bimbard, M. A. Zentile and L. Weller for helpful discussions. 
We acknowledge financial support from EPSRC (grant EP/L023024/1) and Durham University.
The data presented in this paper are available at \url{http://dx.doi.org/10.15128/1c18df763}.
\vfill\eject

\begin{appendix}
\section{Control beam polarization purity measurement}
The transition we use to measure the matrix element is a $\sigma^{-}$ transition and therefore only the fraction of the control beam that has the correct circular polarization will contribute to the EIT.
Since the vapor cell windows are birefringent we need to measure the polarization purity within the vapor cell.
We do this by looking for an EIT resonance which is driven by light of the opposite handedness and measuring the small amount of EIT caused by the impure polarization component.
We also choose to use a transition which has the same dipole moment so that by comparing the extracted Rabi frequencies we have a direct measure of the polarization purity.
We measure a polarization purity of $(99.1\pm0.1)$\%.

\section{Beam profile measurement}\label{beamprofile}
The vapor cell and magnet are removed from the experimental setup and a Thorlabs DC1545M camera is used to record the beam intensity profiles in relative units, $B_{\rm pixel}$.
The absolute calibration of the electric field at each pixel, $E_{\rm pixel}$, is provided by the known pixel size ($d=5.20~\mathrm{\mu m}$ square) and the known total beam power ($P$) through the equation
\begin{equation}
\frac{E_{\rm pixel}}{\sqrt{P}} = \frac{1}{d}\sqrt{\frac{B_{\rm pixel}}{\sum B_{\rm pixel}}}.
\end{equation}

The uncertainty in the axial positioning of the camera with respect to the vapor cell is 1~mm.
The corresponding uncertainty in the matrix element is estimated by calculating the matrix element using beam profiles recorded at axial positions of $\pm1$~mm for the probe beam and $\pm2$~mm for the control beam.
We estimate the errors associated with the probe and control beams to be 0.4\% and 0.5\% respectively.
Therefore by adding these in quadrature the total error associated with the beam profile measurement is 0.7\%.

\section{Systematic line shape effects}\label{lineshapesystematics}
Table \ref{tab:lss} shows a breakdown of potential sources of error in the dipole matrix element measurement caused by systematic changes to the EIT line shape.
\begin{table}
\begin{tabular}{cc}
\hline\hline
Source & Uncertainty (\%) \\
\hline
Control beam intensity distribution & 0.4 \\
EIA modification & negligible \\
Frequency calibration & 0.1 \\
Absolute calibration of transmission & 0.2 \\
Effective probe saturation & 0.08 \\
B field non-uniformity & 0.1 \\
Total & 0.5 \\
\hline\hline
\end{tabular}
\caption{\label{tab:lss} Breakdown of line shape systematics.}
\end{table}
In all cases the uncertainties are estimated by numerically modeling the modified line shapes and then performing a least-squares fit to the three-level EIT model~\cite{Gea-Banacloche1995}.
The following subsections provide details of these systematic changes to the EIT line shape.

\subsection{Control beam intensity distribution}
The three-level EIT line shape~\cite{Gea-Banacloche1995} implicitly assumes that every atom within the probe field experiences the same control-beam electric field.
In practice this can be achieved by using specially engineered diffractive optics~\cite{Gillen-Christandl2015} to create top-hat shaped beams of light.
For simplicity we take the alternate approach of expanding the control beam to approximately double the size of the probe beam.
This approach leads to a lower peak Rabi frequency but has the advantage that the beam profile does not change significantly on propagation through the medium.
We estimate that making the assumption of constant control-beam intensity for our beams leads to a 0.4\% overestimate of the matrix element.

\subsection{EIA modification}\label{EIA}
\begin{figure}
\includegraphics[width=\columnwidth]{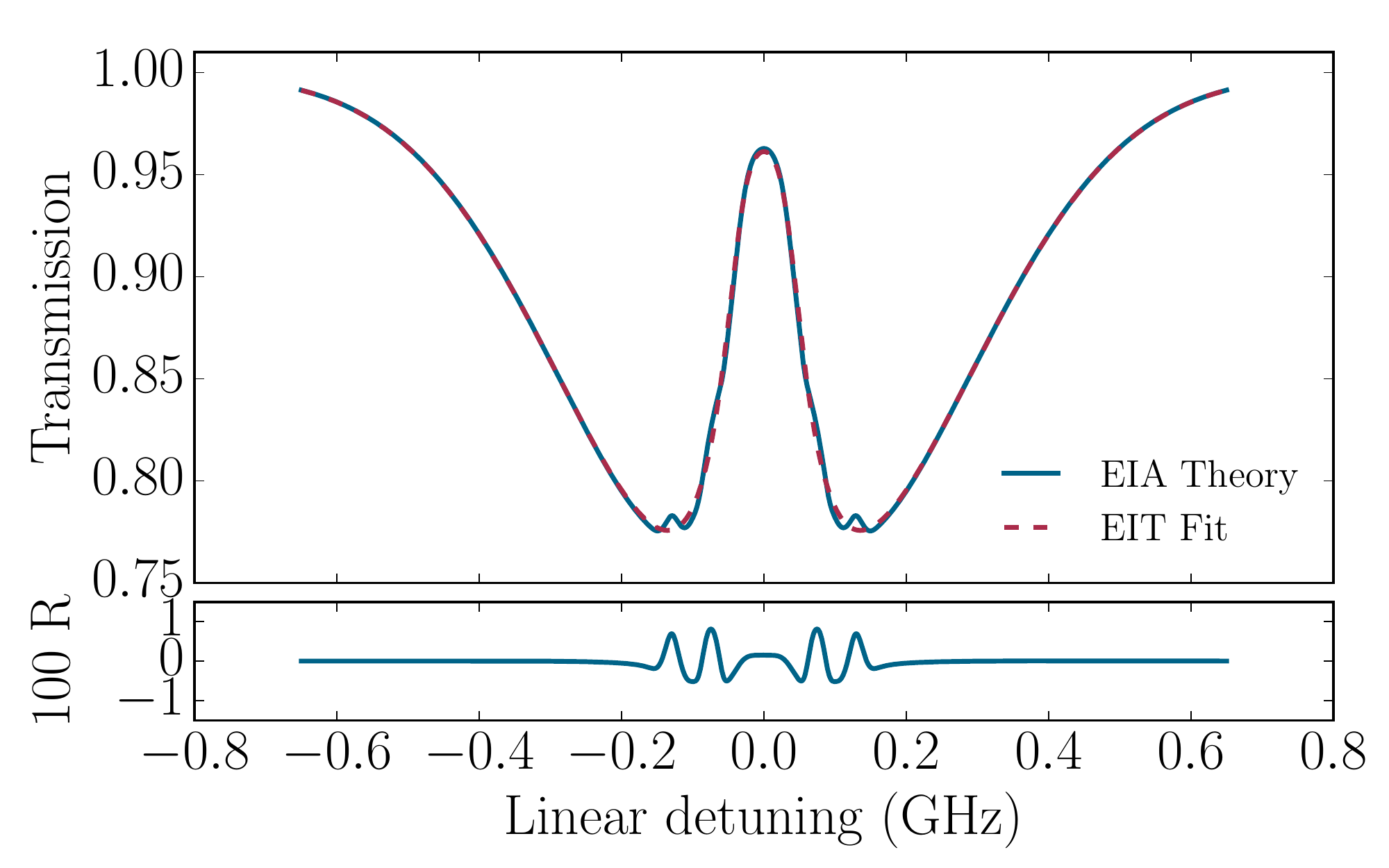}
\caption{\label{fig:eia} Predicted electromagnetically induced absorption (EIA) line shape for a 4\% back reflection of the control beam overlapping with 15\% of the interaction region.
The dashed line shows the fit to this line shape using a three-level EIT line shape and the residual (R) shows the same structure as is found in the experimental data.}
\end{figure}

The vapor cell windows are uncoated and have a reflectivity of approximately 4\% at each surface.
The reflected light from the control beam overlaps with the interaction region and leads to weak electromagnetically induced absorption (EIA) resonances.
This effect has been investigated in detail~\cite{Whiting2015} and its modifying effect on the EIT line shape is well understood.
Figure~\ref{fig:eia} shows the expected EIA line shape assuming a 4\% back reflection overlapping with 15\% of the interaction region (The cell is tilted to minimize this overlap).
The three-level EIT formula [Eq.~(\ref{equation:eit})] is fitted to the modified line shape and the residual shows the same structure that is found in the experimental data.
Although the effect on the line shape is quite noticeable, the effect on the extracted Rabi frequency is small (typically $<0.2$\% change) and the impact on the measurement of the matrix element is even smaller due to the large range of control-beam powers used.

\subsection{B field non-uniformity}
If the vapor cell is not positioned exactly at the center of the magnetic field profile there will be a significant magnetic field gradient across the cell leading to an effective broadening of the lines.
The effect on the line shape can be quite large so care is taken to ensure the cell is positioned correctly.
The uncertainty in the positioning of the cell is $\pm0.5$~mm which leads to a 0.1\% overestimate of the matrix element.

\subsection{Effective probe saturation}
In our experiment the probe-beam Rabi frequency is 0.5~MHz which is significantly smaller than the linewidth of 6~MHz.
As such there is very little power broadening of the spectral lines and the weak probe assumption~\cite{Sherlock2009} is very good.
We estimate that this approximation leads to a 0.08\% underestimate of the matrix element.

\subsection{Frequency calibration}
The experimental spectra are frequency calibrated using an optical cavity with a free spectral range of 375~MHz and an atomic reference based on hyperfine pumping spectroscopy~\cite{Smith2004}.
The uncertainty in the calibration is approximately 0.1\% which directly correlates with the uncertainty in the dipole matrix element.

\subsection{Absolute calibration of transmission}
%The conversion from raw photodiode signal in Volts to absolute transmission of the probe beam is achieved by dividing the raw signal by a polynomial fit to the far off-resonant regions of the spectra, i.e. where we expect the transmission to be unity.
The presence of other nearby absorptive resonances can modify the EIT line shape.
Specifically, the absorption in the wings of the line is increased relative to the line-center value.
Since the nearest absorptive resonances are $\sim1.5$~GHz away, this effect is small and leads to a 0.2\% underestimate of the matrix element.
\end{appendix}

\end{document}